\documentclass[10pt,twocolumn,superscriptaddress,amssymb,amsmath,aps,prd]{revtex4-2}
\usepackage{graphicx}

\usepackage{bbm}

\usepackage[normalem]{ulem}

\usepackage{color}

\newcommand{\cev}[1]{\reflectbox{\ensuremath{\vec{\reflectbox{\ensuremath{#1}}}}}}

\makeatletter
\newcommand{\oset}[3][0ex]{%
  \mathbin{\mathop{#3}\limits^{
    \vbox to#1{\kern-2\ex@
    \hbox{$\scriptstyle#2$}\vss}}}}
\makeatother

\begin{document}


\title{Extending unified gravity to account for graviton--graviton interaction}
\date{August 11, 2025}
\author{Mikko Partanen}
\affiliation{Photonics Group, Department of Electronics and Nanoengineering, 
Aalto University, P.O. Box 13500, 00076 Aalto, Finland}
\author{Jukka Tulkki}
\affiliation{Engineered Nanosystems Group, School of Science, Aalto University, 
P.O. Box 12200, 00076 Aalto, Finland}

\begin{abstract}
Recently, a gauge theory of unified gravity [Rep.~Prog.~Phys.~\textbf{88}, 057802 (2025)] has been developed to extend the Standard Model to include gravity. Here we present unified gravity using the ordinary four-vector and tensor field notation of the Standard Model. The main goal of the present work is to extend the original Minkowski spacetime formulation of the theory to account for graviton--graviton interaction. This is a necessary extension for problems involving interactions between gravitational fields, for example, in the propagation of gravitational waves in external gravitational potentials. The $4\times \mathrm{U(1)}$ gauge invariance of unified gravity is preserved in this extension.
\end{abstract}

\maketitle


\section{Introduction}

In quantum field theory, gauge particles mediate the fundamental forces and interact according to the symmetries of the underlying gauge group \cite{Schwartz2014}. For example, in quantum electrodynamics (QED), the photon is the gauge boson of the Abelian U(1) gauge symmetry. Photons do not interact directly with themselves because they carry no electric charge \cite{Schwartz2014,Peskin2018,Feynman1961}. As a result, there is no fundamental photon--photon interaction in classical electrodynamics or QED. Nevertheless, interactions between photons arise at the loop level of QED, typically via virtual electron–positron pairs, as in light-by-light scattering. These effects are highly suppressed and become relevant only in extreme environments, such as intense electromagnetic fields or high-energy collider processes. In contrast, in non-Abelian gauge theories like quantum chromodynamics (QCD), the gauge particles have mutual interactions \cite{Schwartz2014}. The gluons of QCD self-interact due to their color charge and the non-commutative structure of the SU(3) gauge group.

The situation is fundamentally different for gravitons, the carrier particles of the gravitational force. Gravity, unlike the gauge particles of the Standard Model, couples not to a specific charge but to the total stress-energy-momentum (SEM) tensor, which universally includes contributions from all forms of energy, including the gravitational field itself. As a consequence, gravitons are also sources of gravity, and thus gravitationally self-interacting, but through a different mechanism in comparison with the non-Abelian gauge theories of the Standard Model. Accounting for mutual interactions of gravitons is necessary for the calculation of several phenomena of gravitation, such as the experimentally observable orbital decay of the Hulse--Taylor pulsar \cite{Hulse1975,Freire2024,Will2014}, and gravitational wave propagation in external gravitational fields.

The conventional effective field theory approach to quantum gravity assumes that general relativity (GR) represents a low-energy approximation of a more fundamental quantum field theory \cite{Bambi2023,Casadio2022,Donoghue1994b,Donoghue1999,Donoghue2002,Stelle1977,Schwartz2014,Rocci2024}. In the effective field theory framework, gravity is quantized perturbatively around flat spacetime, with the Einstein–Hilbert action of GR supplemented by an infinite series of higher-dimensional operators encoding quantum corrections. This infinite series of operators means that the effective field theory approach to gravity is not renormalizable in the conventional sense, which would require that all ultraviolet divergences can be reabsorbed into the redefinition of a finite number of parameters of the theory \cite{Schwartz2014,Hooft1971a,Hooft1971b}. Despite being nonrenormalizable in the conventional sense, the effective field theory approach can predict quantum gravity phenomena at energies well below the Planck scale, allowing for systematic computation of quantum gravitational effects such as graviton scattering and loop corrections to classical gravity.

In the conventional effective field theory, gravity arises from the gauge symmetry of diffeomorphism invariance, i.e., general coordinate transformations. In this sense, gravity is treated as a gauge theory of the spacetime metric, which makes it fundamentally distinct from the fundamental interactions of the Standard Model. In contrast, a recently introduced quantum field theory, unified gravity (UG) \cite{Partanen2025a}, describes gravity by the $4\times \mathrm{U(1)}$ tensor gauge field. The $4\times \mathrm{U(1)}$ gauge field is distinct from the metric and appears as an extension of the Standard Model. Furthermore, UG is proposed to be renormalizable in the conventional sense. In the semiclassical limit, UG enables dynamical description of the same phenomena, which are calculated through the metric in GR \cite{Misner1973,Moore2013}. On the relation between UG and GR, we point out that the teleparallel equivalent of GR (TEGR) \cite{Bahamonde2023a,Aldrovandi2012,Maluf2013} results from one particular geometric condition of UG \cite{Partanen2025a}. However, this geometric condition breaks the $4\times \mathrm{U(1)}$ gauge symmetry of UG. Therefore, the pertinent geometric condition makes TEGR fundamentally different from the Minkowski spacetime formulation of UG, which is used in the present work. Accordingly, UG presents a description of gravitational interaction that generally differs from GR already in the classical physics regime.

In the introduction of UG in Ref.~\cite{Partanen2025a}, the graviton--graviton interaction was not included. Consequently, the dynamical equation obtained for the gravitational field in Ref.~\cite{Partanen2025a} is linear, and the resulting description of gravity is not complete regarding all aspects of gravitational interaction. In the present work, we show in detail how the graviton--graviton interaction can be included in UG while preserving the general structure and the Abelian gauge symmetries of UG. The dynamical equation of gravity in UG becomes nonlinear, which is necessary for the correct description of the SEM tensor of gravity acting as a part of the total SEM tensor of all particles and fields of the system.

Instead of the full Standard Model and gravity, here we study the system of the Dirac electron--positron field, the electromagnetic field, and the gravitational field. The original formulation of UG utilized the so-called eight-spinor formalism \cite{Partanen2025a}. In this work, we present UG in the ordinary four-vector and tensor field notation of the Standard Model. Note that in Ref.~\cite{Partanen2025a}, already many key equations were presented both in the eight-spinor representation and in the standard field representation. These representations are mathematically equivalent, i.e., there is a bijective or one-to-one correspondence between them. Consequently, these representations have no influence on the physical interpretation of UG.

The benchmark effects of the gravitational lensing, the perihelion precession of planetary orbits, and gravitational redshift, are investigated using UG in preprints \cite{Partanen2025c,Partanen2025d,Partanen2025e}. Accounting for the graviton--graviton interaction, studied in the present work, does not influence the weak-field limit of these effects.

This work is organized as follows. Section \ref{sec:Lagrangian} presents the generating Lagrangian density of gravity and formulates the gauge theory of UG based on its unitary gauge symmetries. Section \ref{sec:dynamicalequations} derives the dynamical equations of different fields of UG based on the Euler--Lagrange equations. We also present an iterative approach to the solution of the nonlinear field equation of gravity in UG. In Sec.~\ref{sec:TEGR}, we present how TEGR is obtained from UG by a different geometric condition that breaks the gauge symmetry of UG. Finally, conclusions are drawn in Sec.~\ref{sec:conclusion}.

\section{\label{sec:Lagrangian}Lagrangian formulation of the gauge theory of unified gravity}

\subsection{\label{sec:indices}Coordinates and index conventions}

In this work, UG refers to the Minkowski spacetime formulation of the theory \cite{Partanen2025a}. The Latin indices of UG are associated with the Cartesian coordinates $x^a=(ct,x,y,z)$, where $c$ is the speed of light in vacuum and in zero gravitational potential. The coordinates $x^a$ are fixed arbitrarily, and one does not perform any coordinate transformations to them. For the spacetime in UG, one can use arbitrary coordinates of the global Minkowski spacetime, associated with the Greek indices. Any coordinate transformations are applied to $x^\nu$. For simplicity, in this work, we assume Cartesian coordinates $x^\nu=(ct,x,y,z)$, aligned parallel to $x^a$. The choice of using the Cartesian coordinates $x^\nu$ then corresponds to a trivial tetrad, given by the Kronecker delta $\delta_a^\mu$.

The components of the diagonal Minkowski metric tensor $\eta_{\mu\nu}$, in the assumed Cartesian coordinates, are given by $\eta_{00}=1$ and $\eta_{xx}=\eta_{yy}=\eta_{zz}=-1$. In this work, the Einstein summation convention is used for all repeated Greek indices. Below, the repeated Latin indices are, however, not implicitly summed over. The Minkowski metric tensor is the only metric tensor of UG since the gravity gauge-field of UG is not associated with the curved metric as in GR.

\subsection{Generating Lagrangian density of gravity}

The generating Lagrangian density of gravity is the Lagrangian density of the theory at zero gravity gauge field. In the present case, it is then equal to the Lagrangian density of QED. This is shown in detail for the eight-spinor formalism in Sec.~3.7 of Ref.~\cite{Partanen2025a}. In the present work, to obtain complementary insight, we use a reversed approach, where we derive the generating Lagrangian density of gravity starting from the well-known gauge-invariant Lagrangian density of QED, given by
\begin{align}
 \mathcal{L}|_{H=0}
 &=\frac{i\hbar c}{2}\bar{\psi}(\boldsymbol{\gamma}^\mathrm{\nu}\vec{\partial}_\nu-\cev{\partial}_\nu\boldsymbol{\gamma}^\mathrm{\nu})\psi
 -m_\mathrm{e}c^2\bar{\psi}\psi\nonumber\\
 &\hspace{0.4cm}-J_\mathrm{e}^\nu A_\nu
 -\frac{1}{4\mu_0}F_{\mu\nu}F^{\mu\nu}.
 \label{eq:L0_derivation3}
\end{align}
Here  $\hbar$ is the reduced Planck constant, $m_\mathrm{e}$ is the inertial mass of the electron, $\mu_0$ is the permeability of vacuum, $\psi$ is the Dirac spinor, $\bar{\psi}=\psi^\dag\boldsymbol{\gamma}^0$ is the Dirac adjoint, $\boldsymbol{\gamma}^\mu$ are the conventional $4\times 4$ Dirac gamma matrices, $\vec{\partial}_\nu$ and $\cev{\partial}_\nu$ are partial derivatives operating to the right and left, and $J_\mathrm{e}^\nu=q_\mathrm{e}c\bar{\psi}\boldsymbol{\gamma}^\nu\psi$ is the electric four-current density, in which $q_\mathrm{e}$ is the electric charge of the electron or positron. The electromagnetic field-strength tensor, $F_{\mu\nu}$, in Eq.~\eqref{eq:L0_derivation3}, is given in terms of the electromagnetic four-potential $A^\mu$ by the conventional expression as \cite{Landau1989,Jackson1999}
\begin{equation}
 F_{\mu\nu}=\partial_\mu A_\nu-\partial_\nu A_\mu.
\end{equation}

Next, we present the SEM tensor of the Dirac field, denoted by $T_\mathrm{D}^{\mu\nu}$, and the SEM tensor of the electromagnetic gauge field, denoted by $T_\mathrm{em}^{\mu\nu}$, before applying them below. These SEM tensors and their sum, $T_\mathrm{m}^{\mu\nu}$, are given by \cite{Partanen2025a,Partanen2025b}
\begin{align}
 T_\mathrm{m}^{\mu\nu} &=T_\mathrm{D}^{\mu\nu}+T_\mathrm{em}^{\mu\nu},\nonumber\\
 T_\mathrm{D}^{\mu\nu} &=\frac{c}{2}P^{\mu\nu,\rho\sigma}[i\hbar\bar{\psi}(\boldsymbol{\gamma}_\mathrm{\rho}\vec{D}_\sigma-\cev{D}_\rho\boldsymbol{\gamma}_\mathrm{\sigma})\psi-m'_\mathrm{e}c\eta_{\rho\sigma}\bar{\psi}\psi],\nonumber\\
 T_\mathrm{em}^{\mu\nu} &=\frac{1}{\mu_0}\Big(F_{\;\;\rho}^{\mu}F^{\rho\nu}+\frac{1}{4}\eta^{\mu\nu}F_{\rho\sigma}F^{\rho\sigma}\Big)\nonumber\\
 &=\frac{1}{2\mu_0}P^{\mu\nu,\rho\sigma,\eta\lambda}\partial_\rho A_\sigma\partial_\eta A_\lambda.
 \label{eq:semtensors0}
\end{align}
Here $m'_\mathrm{e}$ is the gravitational mass of the electron. Note the factor of $1/2$ in the last form of $T_\mathrm{em}^{\mu\nu}$ in Eq.~\eqref{eq:semtensors0}, which corrects an error in Ref.~\cite{Partanen2025a}, corrected in Ref.~\cite{Partanen2025b}. The right and left electromagnetic gauge-covariant derivatives in $T_\mathrm{D}^{\mu\nu}$ are given by \cite{Landau1982,Peskin2018,Schwartz2014}
\begin{equation}
 \vec{D}_\nu=\vec{\partial}_\nu+i\frac{q_\mathrm{e}}{\hbar}A_\nu,
 \hspace{0.5cm}\cev{D}_\nu=\cev{\partial}_\nu-i\frac{q_\mathrm{e}}{\hbar}A_\nu.
 \label{eq:Da}
\end{equation}
The constant coefficients $P^{\mu\nu,\rho\sigma}$ and $P^{\mu\nu,\rho\sigma,\eta\lambda}$ in Eq.~\eqref{eq:semtensors0} are given by \cite{Partanen2025a,Partanen2025c}
\begin{equation}
 P^{\mu\nu,\rho\sigma}=\frac{1}{2}(\eta^{\mu\sigma}\eta^{\rho\nu}+\eta^{\mu\rho}\eta^{\nu\sigma}-\eta^{\mu\nu}\eta^{\rho\sigma}),
\end{equation}
\begin{align}
 &P^{\mu\nu,\rho\sigma,\eta\lambda}
 =\eta^{\eta\sigma}\eta^{\lambda\mu}\eta^{\nu\rho}-\eta^{\eta\mu}\eta^{\lambda\sigma}\eta^{\nu\rho}-\eta^{\eta\rho}\eta^{\lambda\mu}\eta^{\nu\sigma}\nonumber\\
 &\hspace{0.3cm}+\eta^{\eta\mu}\eta^{\lambda\rho}\eta^{\nu\sigma}-\eta^{\mu\sigma}\eta^{\nu\lambda}\eta^{\rho\eta}+\eta^{\mu\sigma}\eta^{\nu\eta}\eta^{\rho\lambda}+\eta^{\mu\rho}\eta^{\nu\lambda}\eta^{\sigma\eta}\nonumber\\
 &\hspace{0.3cm}-\eta^{\mu\rho}\eta^{\nu\eta}\eta^{\sigma\lambda}-\eta^{\mu\nu}\eta^{\eta\sigma}\eta^{\lambda\rho}+\eta^{\mu\nu}\eta^{\eta\rho}\eta^{\lambda\sigma}.
 \label{eq:P3}
\end{align}

The contraction or trace of the SEM tensor is zero for the SEM tensor of the electromagnetic field in Eq.~\eqref{eq:semtensors0} since photons are massless. For the SEM tensor of the Dirac field in Eq.~\eqref{eq:semtensors0}, we use $P^{\mu\nu,\rho\sigma}\eta_{\mu\nu}=-\eta^{\rho\sigma}$ and $\eta^{\mu\nu}\eta_{\mu\nu}=4$. Consequently, the contractions of the SEM tensors are given by
\begin{align}
 T_\mathrm{em\;\,\nu}^{\;\;\;\,\nu} &=T_\mathrm{D\;\,\nu}^{\;\;\nu}+T_\mathrm{em\;\,\nu}^{\;\;\;\,\nu},
 \hspace{1cm}T_\mathrm{em\;\,\nu}^{\;\;\;\,\nu}=0,\nonumber\\
 T_\mathrm{D\;\,\nu}^{\;\;\nu} &=-\frac{i\hbar c}{2}\bar{\psi}(\boldsymbol{\gamma}^\mathrm{\nu}\vec{D}_\nu-\cev{D}_\nu\boldsymbol{\gamma}^\mathrm{\nu})\psi+2m'_\mathrm{e}c^2\bar{\psi}\psi.
 \label{eq:L0_derivation2}
\end{align}
Using the contractions in Eq.~\eqref{eq:L0_derivation2}, the generating Lagrangian density of gravity in Eq.~\eqref{eq:L0_derivation3} is written as
\begin{align}
 &\mathcal{L}|_{H=0}=-T_\mathrm{m\;\,\nu}^{\;\;\nu}
 +(2m'_\mathrm{e}\!-\!m_\mathrm{e})c^2\bar{\psi}\psi
 -\frac{1}{4\mu_0}F_{\mu\nu}F^{\mu\nu}.
 \label{eq:L0_derivation1}
\end{align}

Following Ref.~\cite{Partanen2025a}, we define a quantity, called the spacetime dimension field. As discussed in detail below, if the generating Lagrangian density of gravity is written using the SEM tensor, the spacetime dimension field obtains a form, given by
\begin{equation}
 I_\mathrm{g}^a=\frac{1}{\sqrt{g_\mathrm{g}}}e^{-ig_\mathrm{g}x_a}.
 \label{eq:Ig}
\end{equation}
Here $g_\mathrm{g}$ is called the scale constant of UG and $x_a=(ct,-x,-y,-z)$. The appearance of $x_a$ in the spacetime dimension field is unconventional in the sence that $x_a$ is not contracted with any four-vector, but the different components of $x_a$ appear individually in different components of $I_\mathrm{g}^a$, which is not a four-vector. As discussed in Se.~\ref{sec:indices}, the Cartesian coordinates $x_a$ are fixed arbitrarily, after which coordinate transformations are not applied to them.

Using the experession of the spacetime dimension field in Eq.~\eqref{eq:Ig}, we obtain an identity, given by
\begin{equation}
 I_\mathrm{g}^{a*}\partial_\nu I_\mathrm{g}^a=-i\delta_a^\mu\eta_{\mu\nu}.
 \label{eq:Ig_identity0}
\end{equation}
As mentioned in Sec.~\ref{sec:indices}, the Einstein summation convention is not applied to Latin indices in this work. Accordingly, there is no implicit summation over $a$ on the left in Eq.~\eqref{eq:Ig_identity0}.


Using Eq.~\eqref{eq:Ig_identity0}, the generating Lagrangian density of gravity in Eq.~\eqref{eq:L0_derivation1} can be written in terms of the spacetime dimension field as
\begin{align}
 &\mathcal{L}|_{H=0}\nonumber\\
 &=-i\sum_a T_\mathrm{m}^{a\nu}I_\mathrm{g}^{a*}\partial_\nu I_\mathrm{g}^a
 +(2m'_\mathrm{e}\!-\!m_\mathrm{e})c^2\bar{\psi}\psi
 -\frac{1}{4\mu_0}F_{\mu\nu}F^{\mu\nu}.
 \label{eq:L0}
\end{align}

The definition of the spacetime dimension field in Eq.~\eqref{eq:Ig} and the generating Lagrangian density of gravity written using the SEM tensor in Eq.~\eqref{eq:L0} could have been used also in the original formulation of UG in Ref.~\cite{Partanen2025a} by representing the SEM tensor and the last two terms of Eq.~\eqref{eq:L0} in terms of the eight-spinor fields.

In the original formulation of UG in Ref.~\cite{Partanen2025a}, the spacetime dimension field had an expression containing $8\times8$ matrices. From the point of view of the present work, this representation is a different but mathematically equivalent definition of the spacetime dimension field that is enabled by the eight-spinor representation in a way that preserves the gauge symmetries of UG.

Above, we have show that the definition of the spacetime dimension field in Eq.~\eqref{eq:Ig} enables transforming the conventional Lagrangian density of QED in Eq.~\eqref{eq:L0_derivation3} into the form in Eq.~\eqref{eq:L0}. The form of the generating Lagrangian density of gravity in Eq.~\eqref{eq:L0} is the starting point for the formulation of the gauge theory associated with the gauge symmetries applied to the spacetime dimension field as discussed below.

\subsection{Four U(1) gauge symmetries of gravity and the conservation law of the SEM tensor}

It is straightforward to observe that the generating Lagrangian density of gravity in Eq.~\eqref{eq:L0} satisfies the following four U(1) symmetries globally:
\begin{equation}
 I_\mathrm{g}^a\rightarrow U_aI_\mathrm{g}^a,\hspace{0.5cm}
 U_a=e^{i\phi_a}.
 \label{eq:transformation}
\end{equation}
Here $\phi_a$ are the symmetry transformation parameters, which are constant for a global symmetry. Comparison with Eq.~(39) of Ref.~\cite{Partanen2025a} shows that the present expression of the spacetime dimension field in Eq.~\eqref{eq:Ig} has the same gauge symmetry properties as the $8\times8$ matrix formulation of the spacetime dimension field in Ref.~\cite{Partanen2025a}.


The infinitesimal variations of the components of the spacetime dimension field with respect to the symmetry transformation parameters $\phi_a$ are given by
\begin{equation}
 \delta I_\mathrm{g}^a=iI_\mathrm{g}^a\delta\phi_a.
\end{equation}

When the generating Lagrangian density of gravity in Eq.~\eqref{eq:L0} is varied with respect to all $\phi_a$ and these variations are summed over, we obtain
\begin{align}
 \delta\mathcal{L}|_{H=0}
 =\frac{1}{g_\mathrm{g}}T_\mathrm{m}^{\mu\nu}\partial_\nu\delta\phi_\mu
 \label{eq:Lagrangiandensityvariation}
\end{align}
This relation is the origin for our definition of the spacetime dimension field and the corresponding representation of the generating Lagrangian density of gravity in Eq.~\eqref{eq:L0}. The variation is analogous to the variation of the Lagrangian density of QED at zero electromagnetic four-potential, given by
$\delta\mathcal{L}_\mathrm{QED}|_{A=0}=-\frac{\hbar}{e}J_\mathrm{e}^\nu\partial_\nu\delta\theta$, where $\theta$ is the U(1) gauge transformation parameter of QED and $J_\mathrm{e}^\nu=q_\mathrm{e}c\bar{\psi}\boldsymbol{\gamma}^\nu\psi$ is the electric four-current density. See the supplementary material of Ref.~\cite{Partanen2025a}.


Next, we consider the derivation of the conservation law of the SEM tensor based on the gauge transformation above. The variation of the action integral with respect to the four U(1) gauge transformation parameters of gravity at zero gravity gauge field becomes
\begin{align}
 &\delta S|_{H=0}=\int\delta\mathcal{L}|_{H=0}d^4x\nonumber\\
 &\!=\int\frac{1}{g_\mathrm{g}}T_\mathrm{m}^{\mu\nu}\partial_\nu\delta\phi_\mu d^4x\nonumber\\
 &\!=-\int\frac{1}{g_\mathrm{g}}(\partial_\nu T_\mathrm{m}^{\mu\nu})\delta\phi_\mu d^4x
 \label{eq:actionvariation}
\end{align}
In the second equality, we have used Eq.~\eqref{eq:Lagrangiandensityvariation}. In the third equality, we have applied partial differentiation and set the total divergence term to zero by assuming that the fields in $T_\mathrm{m}^{\mu\nu}$ vanish at distant boundary.

The last form of equation~\eqref{eq:actionvariation} shows that the variation of the action integral vanishes for arbitrary $\delta\phi_\mu$ when
\begin{equation}
 \partial_\nu T_\mathrm{m}^{\mu\nu}=0.
 \label{eq:Tmconservation}
\end{equation}
This is the well-known conservation law of the SEM tensor in Cartesian coordinates at zero gravitational field.


\subsection{Gauge-covariant derivative}

Next, we formulate the locally gauge-invariant theory by the introduction of the gravity gauge field through the gauge-covariant derivative as elaborated in Sec.~5.1 of Ref.~\cite{Partanen2025a}. The generating Lagrangian density of gravity in Eq.~\eqref{eq:L0} is globally gauge invariant in the symmetry transformation of Eq.~\eqref{eq:transformation}. In the global symmetry transformation, the values of $\phi_a$ are constant. To promote the global symmetry to a local symmetry, we allow $\phi_a$ to depend on the spacetime coordinates $x^\mu$. Following conventional gauge theory \cite{Peskin2018,Weinberg1996}, the generating Lagrangian density of gravity in Eq.~\eqref{eq:L0} can be made locally gauge invariant in the symmetry transformation of Eq.~\eqref{eq:transformation} when we generalize the partial derivative that acts on $I_\mathrm{g}^a$ into a gauge-covariant derivative $\mathcal{D}_\nu$, defined as
\begin{equation}
 \mathcal{D}_\nu I_\mathrm{g}^a=(\partial_\nu-ig'_\mathrm{g}H_{a\nu})I_\mathrm{g}^a.
 \label{eq:covariantderivative}
\end{equation}
Here $g'_\mathrm{g}$ is the coupling constant of UG and $H_{a\nu}$ is the gravity gauge field \cite{Partanen2025a}. The gauge transformation of $H_{a\nu}$ is given by
\begin{equation}
 H_{a\nu}\rightarrow H_{a\nu}+\frac{1}{g'_\mathrm{g}}\partial_\nu\phi_a.
 \label{eq:Hanutransformation}
\end{equation}
This transformation of $H_{a\nu}$ makes $\mathcal{D}_\nu I_\mathrm{g}^a$ gauge invariant when $I_\mathrm{g}^a$ is transformed according to Eq.~\eqref{eq:transformation}.
Substituting $\mathcal{D}_\nu I_\mathrm{g}^a$ in Eq.~\eqref{eq:covariantderivative} in place of $\partial_\nu I_\mathrm{g}^a$ in the generating Lagrangian density of gravity in Eq.~\eqref{eq:L0} makes the Lagrangian density locally gauge invariant with respect to the gauge transformations in Eqs.~\eqref{eq:transformation} and \eqref{eq:Hanutransformation}.

%
%
%

\subsection{Gravity gauge field strength tensor}

To construct the full gauge-invariant Lagrangian density, we need to add a term that involves only the gauge field $H_{a\nu}$, and this term must itself be gauge invariant. Gauge theory provides a well-defined way to do this by utilizing the commutator of the gauge-covariant derivatives \cite{Peskin2018,Weinberg1996}. Then, we obtain a unique expression for the antisymmetric gravity gauge-field-strength tensor $H_{a\mu\nu}$ as
\begin{equation}
 \begin{array}{c}
 [\mathcal{D}_\mu,\mathcal{D}_\nu]I_\mathrm{g}^a=-ig'_\mathrm{g}H_{a\mu\nu}I_\mathrm{g}^a,\\[10pt]
 H_{a\mu\nu}=\partial_\mu H_{a\nu}-\partial_\nu H_{a\mu},\\[10pt]
 H_{\rho\mu\nu}=\delta_\rho^a H_{a\mu\nu}.
 \end{array}
 \label{eq:Hdefinition}
\end{equation}
In the last form of Eq.~\eqref{eq:Hdefinition}, we use the trivial tetrad related to our choice of using the Cartesian coordinates $x^\nu$ aligned parallel to $x^a$. Since UG is an Abelian gauge theory, the gauge-field-strength tensor $H_{a\mu\nu}$ is invariant in the gauge transformation, given in Eq.~\eqref{eq:Hanutransformation}.

\subsection{Lagrangian density of the gravity gauge field}

In analogy with the gauge theories in the Standard Model \cite{Peskin2018}, the Lagrangian density for the gravity gauge field strength is not uniquely fixed by gauge invariance alone. In the Standard Model, the form of the gauge field Lagrangian is further constrained by requirements such as parity and time-reversal symmetries, as well as renormalizability \cite{Peskin2018}. As detailed in Ref.~\cite{Partanen2025a}, the Lagrangian density of the gravity gauge field strength is given by
\begin{equation}
 \mathcal{L}_\mathrm{g,kin}=\frac{1}{4\kappa}H_{\rho\mu\nu}S^{\rho\mu\nu}.
 \label{eq:Lgkin}
\end{equation}
Here $S^{\rho\mu\nu}$ is the superpotential and $\kappa=8\pi G/c^4$ is Einstein's constant, in which $G$ is the gravitational constant. The prefactor of equation~\eqref{eq:Lgkin} has been determined by comparison of the weak field limit of UG with Newton's law of gravitation \cite{Partanen2025a}. The superpotential is given by
\begin{align}
 S^{\rho\mu\nu} &=\frac{1}{2}(H^{\nu\mu\rho}+H^{\mu\rho\nu}-H^{\rho\nu\mu})+\eta^{\rho\mu}H_{\;\;\;\;\,\sigma}^{\sigma\nu}-\eta^{\rho\nu}H_{\;\;\;\;\,\sigma}^{\sigma\mu}\nonumber\\
 &=C^{\rho\mu\nu,\alpha\beta\gamma}H_{\alpha\beta\gamma}.
 \label{eq:superpotential}
\end{align}
Here we have defined the coefficients $C^{\rho\mu\nu,\alpha\beta\gamma}$ as
\begin{align}
 C^{\rho\mu\nu,\alpha\beta\gamma} &=\frac{1}{2}(\eta^{\nu\alpha}\eta^{\mu\beta}\eta^{\rho\gamma}+\eta^{\mu\alpha}\eta^{\rho\beta}\eta^{\nu\gamma}-\eta^{\rho\alpha}\eta^{\nu\beta}\eta^{\mu\gamma})\nonumber\\
 &\hspace{0.4cm}+\eta^{\rho\mu}\eta^{\alpha\gamma}\eta^{\nu\beta}-\eta^{\rho\nu}\eta^{\alpha\gamma}\eta^{\mu\beta}.
\end{align}

\subsection{Accounting for graviton--graviton interaction}

Above, we have followed the original formulation of UG in Ref.~\cite{Partanen2025a}, but presented it in the standard field formulation. Next, we present an extension that accounts for the graviton--graviton interaction. The SEM tensor of the gravity gauge field did not appear as a source term of the gravitational field in the Minkowski spacetime formulation in Ref.~\cite{Partanen2025a}. The gauge theory gives us the freedom to introduce an additional gauge-invariant Lagrangian density term, which depends on the spacetime dimension field and the gravity gauge field strength, given by
\begin{equation}
 \mathcal{L}_\mathrm{gg,int}=-i\sum_a T_\mathrm{g}^{a\nu}I_\mathrm{g}^{a*}\mathcal{D}_\nu I_\mathrm{g}^a.
 \label{eq:Lggint}
\end{equation}
As will be explained in detail below, adding this term enables us to establish a more general conservation law of the SEM tensor in comparison with the conservation law at zero gravitational field in Eq.~\eqref{eq:Tmconservation}. In Eq.~\eqref{eq:Lggint}, $T_\mathrm{g}^{a\nu}$ is the gauge-invariant, symmetric SEM tensor of the gravity gauge field, written as
\begin{align}
 T_\mathrm{g}^{\mu\nu} &=\frac{1}{2\kappa}\Big(H^{\;\,\;\;\mu}_{\rho\sigma}S^{\rho\sigma\nu}+H^{\;\,\;\;\nu}_{\rho\sigma}S^{\rho\sigma\mu}-\frac{1}{2}\eta^{\mu\nu}H_{\rho\sigma\lambda}S^{\rho\sigma\lambda}\Big)\nonumber\\
 &=\frac{1}{2\kappa}P^{\mu\nu,\rho\sigma\lambda,\alpha\beta\gamma}\partial_\rho H_{\sigma\lambda}\partial_\alpha H_{\beta\gamma}.
 \label{eq:Tg}
\end{align}
This SEM tensor is formed from the second-order terms of the derivatives of the gravity gauge field. It is traceless since gravitons are massless. The constant coefficients $P^{\mu\nu,\rho\sigma\lambda,\alpha\beta\gamma}$ in Eq.~\eqref{eq:Tg} are given by
\begin{align}
 P^{\mu\nu,\sigma\rho\lambda,\beta\alpha\gamma}
 &=D^{\lambda\mu,\rho\sigma\nu,\alpha\beta\gamma}-D^{\lambda\mu,\rho\sigma\nu,\alpha\gamma\beta}\nonumber\\
 &\hspace{0.4cm}-D^{\sigma\mu,\rho\lambda\nu,\alpha\beta\gamma}+D^{\sigma\mu,\rho\lambda\nu,\alpha\gamma\beta},\nonumber\\
 D^{\lambda\mu,\rho\sigma\nu,\alpha\beta\gamma} &=\eta^{\lambda\mu}C^{\rho\sigma\nu,\alpha\beta\gamma}+\eta^{\lambda\nu}C^{\rho\sigma\mu,\alpha\beta\gamma}\nonumber\\
 &\hspace{0.4cm}-\frac{1}{2}\eta^{\mu\nu}C^{\rho\sigma\lambda,\alpha\beta\gamma}.
\end{align}

The main motivation for the definition of $\mathcal{L}_\mathrm{gg,int}$ in Eq.~\eqref{eq:Lggint}, as shown in detail in Sec.~\ref{sec:conservationlaw} below, is that it leads to the total SEM tensor $T^{\mu\nu}$ of the Dirac electron--positron field, the electromagnetic gauge field, and the gravity gauge field, given by the sum of the SEM tensors in Eqs.~\eqref{eq:semtensors0} and \eqref{eq:Tg} as
\begin{equation}
 T^{\mu\nu}=T_\mathrm{m}^{\mu\nu}+T_\mathrm{g}^{\mu\nu}.
 \label{eq:Ttot}
\end{equation}

\subsection{Gauge-invariant Lagrangian density}
Using the gauge-covariant derivative in Eq.~\eqref{eq:covariantderivative} and adding the terms associated with the gravity gauge field in Eqs.~\eqref{eq:Lgkin} and \eqref{eq:Lggint} to the generating Lagrangian density of gravity in Eq.~\eqref{eq:L0}, we obtain the locally gauge-invariant Lagrangian density. This Lagrangian density satisfies locally the electromagnetic [U(1)] gauge-invariance and the gravity [4$\times$U(1)] gauge-invariance, and it is given by
\begin{align}
 \mathcal{L} &=-i\sum_a T^{a\nu}I_\mathrm{g}^{a*}\mathcal{D}_\nu I_\mathrm{g}^a
 +(2m'_\mathrm{e}-m_\mathrm{e})c^2\bar{\psi}\psi\nonumber\\
 &\hspace{0.4cm}-\frac{1}{4\mu_0}F_{\mu\nu}F^{\mu\nu}
 +\frac{1}{4\kappa}H_{\rho\mu\nu}S^{\rho\mu\nu}.
 \label{eq:LLL}
\end{align}
The Lagrangian density in Eq.~\eqref{eq:LLL} now includes the graviton--graviton interaction through $\mathcal{L}_\mathrm{gg,int}$, which was not accounted for in Ref.~\cite{Partanen2025a}. This term makes the total SEM tensor $T^{a\nu}$ to appear in the gauge-invariant Lagrangian density in Eq.~\eqref{eq:LLL}.

\subsection{\label{sec:conservationlaw}Generalized conservation law of the SEM tensor}

Here we derive the generalized conservation law of the SEM tensor in the presence of gravitational interaction. In analogy with Eq.~\eqref{eq:Lagrangiandensityvariation}, we can calculate the variation of the gauge-invariant Lagrangian density, given in Eq.~\eqref{eq:LLL}, with respect to the gauge symmetry transformation parameters in a nonzero gravity gauge field. Applying the gauge transformation to the spacetime dimension field and preserving the gravity gauge field fixed in the variation, we obtain
\begin{align}
 \delta\mathcal{L}|_{H}
 =\frac{1}{g_\mathrm{g}}T^{\mu\nu}\partial_\nu\delta\phi_\mu.
 \label{eq:Lagrangiandensityvariation2}
\end{align}
The variation in Eq.~\eqref{eq:Lagrangiandensityvariation2} depends on the gravity gauge field through $T_\mathrm{g}^{\mu\nu}$, which is part of $T^{\mu\nu}$. This is associated with the energy content of the gravitational field itself. In contrast, the variation of the Lagrangian density of QED at constant electromagnetic four-potential, given by
$\delta\mathcal{L}_\mathrm{QED}|_{A}=-\frac{\hbar}{e}J_\mathrm{e}^\nu\partial_\nu\delta\theta$, \emph{does not depend} on the value of the four-potential. This is because the four-potential does not have electric charge so that it could contribute to the electric four-current density.

The variation of the action integral corresponding to the variation in Eq.~\eqref{eq:Lagrangiandensityvariation2} becomes, in analogy with Eq.~\eqref{eq:actionvariation}, zero when the total SEM tensor in Eq.~\eqref{eq:Ttot} satisfies the conservation law, given by
\begin{equation}
 \partial_\nu T^{\mu\nu}=0.
 \label{eq:conservationT}
\end{equation}
This is the generalization of the conservation law in Eq.~\eqref{eq:Tmconservation} for nonzero gravitational fields. The conservation law in Eq.~\eqref{eq:conservationT} means that the energies, momenta, and angular momenta of the Dirac and electromagnetic fields in $T_\mathrm{m}^{\mu\nu}$ are no longer conserved but they can be converted into energy, momentum, and angular momentum of the gravitational field, described through $T_\mathrm{g}^{\mu\nu}$, which is part of $T^{\mu\nu}$.

\subsection{Faddeev--Popov gauge-fixed Lagrangian density}

It is well known that the formulation of a gauge field theory requires fixing the gauge \cite{Peskin2018,Schwartz2014}. This necessity arises because gauge theories describe physical field configurations by equivalence classes defined under gauge transformations. These classes reflect the presence of redundant degrees of freedom in the gauge fields. Therefore, one must eliminate this redundancy through a process called gauge fixing. In UG, we follow the well-known Faddeev--Popov method \cite{Peskin2018,Schwartz2014,Faddeev1967}. The Faddeev--Popov gauge-fixed Lagrangian density of UG is given by \cite{Partanen2025a}
\begin{equation}
 \mathcal{L}_\mathrm{FP}=\mathcal{L}+\mathcal{L}_\mathrm{em,gf}+\mathcal{L}_\mathrm{em,ghost}+\mathcal{L}_\mathrm{g,gf}+\mathcal{L}_\mathrm{g,ghost}.
 \label{eq:LFP}
\end{equation}
Here the electromagnetic and gravity gauge-fixing and ghost Lagrangian densities are given by \cite{Partanen2025a}
\begin{align}
 \mathcal{L}_\mathrm{em,gf} &
 =-\frac{1}{2\mu_0\xi_\mathrm{e}}[C_\mathrm{em}(A)]^2=-\frac{1}{2\mu_0\xi_\mathrm{e}}(\partial_{\nu}A^{\nu})^2,
 \label{eq:QEDfixing}
\end{align}
\begin{equation}
 \mathcal{L}_\mathrm{em,ghost} =\hbar c\bar{c}_\mathrm{em}\partial^2 c_\mathrm{em},
 \label{eq:Lemghost}
\end{equation}
\begin{align}
 \mathcal{L}_\mathrm{g,gf}
 &=\frac{1}{4\kappa\xi_\mathrm{g}}C_\mathrm{g}^\mu(H) C_{\mathrm{g}\mu}(H)\nonumber\\
 &=\frac{1}{\kappa\xi_\mathrm{g}}\eta_{\gamma\delta}P^{\alpha\beta,\lambda\gamma}P^{\rho\sigma,\eta\delta}\partial_\lambda H_{\alpha\beta}\partial_\eta H_{\rho\sigma},
 \label{eq:Lggf}
\end{align}
\begin{equation}
 \mathcal{L}_\mathrm{g,ghost}=-\hbar c\bar{c}_\mathrm{g}\partial^2 c_\mathrm{g}.
 \label{eq:Lgghost}
\end{equation}
Here $\xi_\mathrm{e}$ and $\xi_\mathrm{g}$ are the electromagnetic and gravity gauge-fixing parameters, respectively.

\subsection{BRST invariance}
The Faddeev--Popov Lagrangian density of UG in Eq.~\eqref{eq:LFP} exhibits an exact global symmetry known as BRST invariance in analogy with the gauge theories of the Standard Model \cite{Becchi1976,Iofa1976,Peskin2018,Schwartz2014,Fuster2005}. In the BRST formalism, the local U(1) gauge parameter of QED is replaced by $\theta = \theta' c_\mathrm{em}$, where $\theta'$ is a constant anticommuting Grassmann number with $\theta'^2 = 0$. Similarly, the four gravity U(1) gauge parameters are replaced by $\phi_\mu = \phi' c_{\mathrm{g} \mu}$, where $\phi'$ is a constant Grasmann number with $\phi'^2 = 0$. Under these substitutions, the Lagrangian density in Eq.~~\eqref{eq:LFP} remains invariant under the BRST transformations for electromagnetism, given by \cite{Schwartz2014,Partanen2025a}
\begin{align}
 \psi &\rightarrow e^{i\theta' c_\mathrm{em}Q}\psi,\nonumber\\
 A_\nu &\rightarrow A_\nu-\frac{\hbar}{e}\theta'\partial_\nu c_\mathrm{em},\nonumber\\
 \bar{c}_\mathrm{em} &\rightarrow\bar{c}_\mathrm{em}-\frac{1}{\mu_0ce\xi_\mathrm{e}}\theta'C_\mathrm{em}(A),\nonumber\\
 c_\mathrm{em} &\rightarrow c_\mathrm{em},
\end{align}
and under the BRST transformations associated with gravity, given by \cite{Partanen2025a}
\begin{align}
 I_\mathrm{g}^a&\rightarrow e^{i\phi'c_{\mathrm{g}a}}I_\mathrm{g}^a,\nonumber\\
 H_{a\nu} &\rightarrow H_{a\nu}+\frac{1}{g'_\mathrm{g}}\phi'\partial_\nu c_{\mathrm{g}a},\nonumber\\
 \bar{c}_\mathrm{g}^a &\rightarrow\bar{c}_\mathrm{g}^a-\frac{1}{\kappa\hbar cg'_\mathrm{g}\xi_\mathrm{g}}\phi'C_\mathrm{g}^a(H),\nonumber\\
 c_\mathrm{g}^a &\rightarrow c_\mathrm{g}^a.
\end{align}

BRST symmetry is known to hold at all loop orders in the path integral formalism \cite{Schwartz2014}. Its presence in UG strongly indicates that the theory is renormalizable, like the gauge theories of the Standard Model. This is further supported by the successful one-loop renormalization of UG in Ref.~\cite{Partanen2025a}. Unlike conventional gravity theories, where gauge generators are field-dependent, UG features constant generators, allowing BRST symmetry to apply directly, without requiring the more general Batalin--Vilkovisky formalism \cite{Batalin1981,Batalin1983,Weinberg1996,Costello2011,Henneaux1992,Gomis1995}.

\subsection{Reduced form of the Lagrangian density}

Next, we present the reduced form of the Lagrangian density of UG by writing the spacetime dimension field explicitly using Eq.~\eqref{eq:Ig}. Furthermore, in this Lagrangian density, we drop out the ghost-field terms of the Lagrangian density, which do not participate in the dynamics of fields in Abelian gauge theories, such as UG. By applying the expression of the spacetime dimension field in Eq.~\eqref{eq:Ig}, we obtain an identity
\begin{equation}
 I_\mathrm{g}^{a*}\mathcal{D}_\nu I_\mathrm{g}^a=-i\delta_a^\mu\Big(\eta_{\mu\nu}+\frac{g'_\mathrm{g}}{g_\mathrm{g}}H_{\mu\nu}\Big).
 \label{eq:Ig_identity}
\end{equation}
Using the identity in Eq.~\eqref{eq:Ig_identity} in the Faddeev--Popov Lagrangian density in Eq.~\eqref{eq:LFP} and dropping out the ghost-field terms, the reduced form of the Lagrangian density of UG becomes
\begin{align}
 \mathcal{L}_\mathrm{UG}&=\frac{i\hbar c}{2}\bar{\psi}(\boldsymbol{\gamma}^\nu\vec{\partial}_\nu-\cev{\partial}_\nu\boldsymbol{\gamma}^\nu)\psi-m_\mathrm{e}c^2\bar{\psi}\psi
 -\frac{1}{4\mu_0}F_{\mu\nu}F^{\mu\nu}\nonumber\\
 &+\frac{1}{4\kappa}H_{\rho\mu\nu}S^{\rho\mu\nu}
 \!-\!J_\mathrm{e}^\nu A_\nu\!-\!\frac{g'_\mathrm{g}}{g_\mathrm{g}}T^{\mu\nu}H_{\mu\nu}
 \!-\!\frac{1}{2\mu_0\xi_\mathrm{e}}(\partial_{\nu}A^{\nu})^2\nonumber\\
 &+\frac{1}{\kappa\xi_\mathrm{g}}\eta_{\gamma\delta}P^{\alpha\beta,\lambda\gamma}P^{\rho\sigma,\eta\delta}\partial_\lambda H_{\alpha\beta}\partial_\eta H_{\rho\sigma}.
 \label{eq:LUG}
\end{align}

In comparison with the reduced form of the Lagrangian density of UG in Eqs.~(108) and (109) of Ref.~\cite{Partanen2025a}, the only change in Eq.~\eqref{eq:LUG} is the appearance of the total SEM tensor $T^{\mu\nu}$, given in \eqref{eq:Ttot}, in place of the SEM tensor $T_\mathrm{m}^{\mu\nu}$ of the Dirac and electromagnetic fields. Consequently, the pertinent term depends on the third power of the gravity gauge field. As shown in Sec.~\ref{sec:dynamicalequations}, the dynamical equation of gravity in UG obtains terms depending on the second power of the gravity gauge field. Thus, the dynamical equation of gravity becomes nonlinear.

\subsection{Equivalence principle of unified gravity}

The Lagrangian density of UG has been formulated above using the gravitational mass $m'_\mathrm{e}$, the inertial mass $m_\mathrm{e}$, the scale constant $g_\mathrm{g}$, and the coupling constant $g'_\mathrm{g}$. The equivalence principles of UG between these quantities are given by \cite{Partanen2025a}
\begin{equation}
 m'_\mathrm{e}=m_\mathrm{e},\hspace{1cm}
 g'_\mathrm{g}=g_\mathrm{g}.
 \label{eq:equivalence}
\end{equation}
The first relation in Eq.~\eqref{eq:equivalence} is called the equivalence principle of mass, and the second relation is called the equivalence principle of scale. The equivalence principles in Eq.~\eqref{eq:equivalence} apply to the parameters of the classical theory and to the renormalized values of the pertinent quantities in the quantum field theory \cite{Partanen2025a}.

\section{\label{sec:dynamicalequations}Dynamical equations}

Dynamical equations of all fields in the Lagrangian density can be straightforwardly derived through the Euler--Lagrange equations. Here we use the reduced form of the Lagrangian density of UG in Eq.~\eqref{eq:LUG}. In the derivation of the dynamical equations, we also apply the equivalence principles of UG, given in Eq.~\eqref{eq:equivalence}.

\subsection{Dynamical equation of the gravitational field}
Using the Euler--Lagrange equations, written for $H_{\mu\nu}$, the dynamical equation of the gravity gauge field, in the harmonic gauge with $\xi_\mathrm{g}=1$, becomes
\begin{equation}
 P^{\mu\nu,\rho\sigma}\partial^2H_{\rho\sigma}-P^{\sigma\lambda,\rho\mu\nu,\alpha\beta\gamma}\partial_\rho(H_{\sigma\lambda}\partial_\alpha H_{\beta\gamma})=-\kappa T^{\mu\nu}.
 \label{eq:UGgravitygeneral}
\end{equation}
This equation is different from the dynamical equation of gravity obtained in Ref.~\cite{Partanen2025a} by the nonlinear terms, i.e., the second term on the left and the SEM tensor of gravity, $T_\mathrm{g}^{\mu\nu}$, which is part of the total SEM tensor $T^{\mu\nu}$ on the right. Both these terms follow from the Lagrangian density term $\mathcal{L}_\mathrm{gg,int}$ in Eq.~\eqref{eq:Lggint}, which was not included in Ref.~\cite{Partanen2025a}.

\subsection{Dynamical equation of the electromagnetic field}
Using the Euler--Lagrange equation, written for $A_\sigma$, we obtain the dynamical equation of the electromagnetic four-potential, in the Feynman gauge with $\xi_\mathrm{e}=1$, given by \cite{Partanen2025a,Partanen2025b}
\begin{align}
 \partial^2A^\sigma+P^{\mu\nu,\rho\sigma,\eta\lambda}\partial_\rho(H_{\mu\nu}\partial_\eta A_\lambda) &=\mu_0J_\mathrm{e,tot}^\sigma.
 \label{eq:MaxwellUG}
\end{align}
Here $J_\mathrm{e,tot}^\mu$ is the total electric four-current density in the presence of the gravity gauge field, given by
\begin{equation}
 J_\mathrm{e,tot}^\mu=J_\mathrm{e}^\rho-P^{\mu\nu,\rho\sigma}J_\mathrm{e\sigma}H_{\mu\nu}.
 \label{eq:Jetot}
\end{equation}
In the presence of gravitational interaction, the total electric four-current density $J_\mathrm{e,tot}^\mu$ in Eq.~\eqref{eq:Jetot} satisfies the conservation law, given by $\partial_\nu J_\mathrm{e,tot}^\nu=0$. The conservation law does not hold for $J_\mathrm{e}^\nu=q_\mathrm{e}c\bar{\psi}\boldsymbol{\gamma}^\nu\psi$. This is discussed in Ref.~\cite{Partanen2025e}. We conclude that the graviton--graviton interaction does no influence the form of the terms in Eq.~\eqref{eq:MaxwellUG}.

\subsection{Dynamical equation of the Dirac field}
Using the Euler--Lagrange equation, written for $\bar{\psi}$, the dynamical equation of the Dirac field becomes \cite{Partanen2025a}
\begin{align}
 &i\hbar c\boldsymbol{\gamma}^\rho\vec{\partial}_\rho\psi\!-\!m_\mathrm{e}c^2\psi
 \!=q_\mathrm{e}c\boldsymbol{\gamma}^\rho\psi A_\rho
 +P^{\mu\nu,\rho\sigma}\Big(i\hbar c\boldsymbol{\gamma}_\sigma\vec{\partial}_\rho\psi\nonumber\\
 &\hspace{0.4cm}-\frac{m_\mathrm{e}c^2}{2}\eta_{\rho\sigma}\psi+\frac{i\hbar c}{2}\boldsymbol{\gamma}_\sigma\psi\vec{\partial}_\rho
 -q_\mathrm{e}c\boldsymbol{\gamma}_\sigma\psi A_\rho
 \Big)H_{\mu\nu}.
 \label{eq:DiracUG}
\end{align}
The graviton--graviton interaction does no influence the form of the terms in Eq.~\eqref{eq:DiracUG}.

\subsection{Linearized dynamical equation of gravity and iterative approach to the nonlinear equation}

\subsubsection{Linearized equation of gravity}

Assuming the gravity gauge field small, the nonlinear terms associated with the graviton--graviton interaction in Eq.~\eqref{eq:UGgravitygeneral} can be neglected. Then, the linearization of Eq.~\eqref{eq:UGgravitygeneral} leads to the inhomogeneous wave equation, given by \cite{Partanen2025a,Partanen2025c}
\begin{equation}
 P^{\mu\nu,\rho\sigma}\partial^2H_{\rho\sigma}^{(0)}=-\kappa T_\mathrm{m}^{\mu\nu}.
 \label{eq:UGgravity}
\end{equation}
Here $H_{\rho\sigma}^{(0)}$ denotes the gravity gauge field solution of the linearized equation. The source term on the right is the SEM tensor $T_\mathrm{m}^{\mu\nu}$ of the Dirac and electromagnetic fields, which neglects the SEM tensor of the gravitational field, $T_\mathrm{g}^{\mu\nu}$ in Eq.~\eqref{eq:Tg}. The linearized equation in Eq.~\eqref{eq:UGgravity} is equivalent to the field equation of gravity in the absence of graviton--graviton interaction as derived in Ref.~\cite{Partanen2025a}.

As the solution of Eq.~\eqref{eq:UGgravity}, the gravity gauge field is obtained from the source term $T_\mathrm{m}^{\mu\nu}$ through the retarded Green's function of the wave equation \cite{Poisson2014,Misner1973}. We also use the identity $P_{\mu\nu,\alpha\beta}P^{\alpha\beta,\rho\sigma}=I_{\mu\nu}^{\rho\sigma}$, where $I_{\mu\nu}^{\rho\sigma}=\frac{1}{2}(\delta_\mu^\rho\delta_\nu^\sigma+\delta_\nu^\rho\delta_\mu^\sigma)$ is the identity tensor \cite{Partanen2025a}. Then, the solution for $H_{\mu\nu}^{(0)}$ is given by
\begin{equation}
 H_{\mu\nu}^{(0)}=-\frac{\kappa P_{\mu\nu,\rho\sigma}}{4\pi}\int\frac{T_\mathrm{m}^{\rho\sigma}(t_\mathrm{r},\mathbf{r}')}{|\mathbf{r}-\mathbf{r}'|}d^3r'.
 \label{eq:solution}
\end{equation}
Here $t_\mathrm{r}=t-\frac{1}{c}|\mathbf{r}-\mathbf{r}'|$ is the retarded time, defined in terms of coordinates of the global Minkowski frame.

\subsubsection{Iterative approach to the nonlinear equation of gravity}

Next, we construct an iterative approach for the solution of the nonlinear dynamical equation of the gravity gauge field in Eq.~\eqref{eq:UGgravitygeneral}. The zeroth-order solution, denoted by $H_{\rho\sigma}^{(0)}$, is taken from the linearized equation of gravity in Eq.~\eqref{eq:UGgravity}. The first-order solution, denoted by $H_{\rho\sigma}^{(1)}$, is obtained by approximating the nonlinear terms of the field equation in Eq.~\eqref{eq:UGgravitygeneral} by their values calculated using the zeroth-order solution $H_{\rho\sigma}^{(0)}$. This iterative approach can be continued to arbitrary order $n$ by writing the approximation of Eq.~\eqref{eq:UGgravitygeneral} as
\begin{align}
 &P^{\mu\nu,\rho\sigma}\partial^2H_{\rho\sigma}^{(n)}-P^{\sigma\lambda,\rho\mu\nu,\alpha\beta\gamma}\partial_\rho(H_{\sigma\lambda}^{(n-1)}\partial_\alpha H_{\beta\gamma}^{(n-1)})\nonumber\\
 &=-\kappa(T_\mathrm{m}^{\mu\nu}+T_\mathrm{g}^{(n-1)\mu\nu}),\hspace{0.5cm}\text{for }n\ge1.
 \label{eq:UGgravityn}
\end{align}
Here $T_\mathrm{g}^{(n-1)\mu\nu}$ is the approximation of the SEM tensor of gravity, calculated using Eq.~\eqref{eq:Tg} for $H_{\rho\sigma}^{(n-1)}$.

Equation \eqref{eq:UGgravityn} is an inhomogeneous wave equation for $H_{\rho\sigma}^{(n)}$. Thus, it is straightforward to solve using the retarded Green's function of the wave equation in analogy to Eqs.~\eqref{eq:UGgravity} and \eqref{eq:solution} as
\begin{align}
 H_{\mu\nu}^{(n)} &=-\frac{\kappa P_{\mu\nu,\rho\sigma}}{4\pi}\!\int\!\frac{1}{|\mathbf{r}\!-\!\mathbf{r}'|}\Big[T_\mathrm{m}^{\rho\sigma}(t_\mathrm{r},\mathbf{r}')\!+\!T_\mathrm{g}^{(n-1)\rho\sigma}(t_\mathrm{r},\mathbf{r}')\nonumber\\
 &\hspace{0.4cm}-\frac{1}{\kappa}P^{\eta\lambda,\delta\rho\sigma,\alpha\beta\gamma}\partial_\delta(H_{\eta\lambda}^{(n-1)}\partial_\alpha H_{\beta\gamma}^{(n-1)})\Big]d^3r'.
 \label{eq:solutioniterative}
\end{align}
For weak fields, the nonlinear terms of Eqs.~\eqref{eq:UGgravityn} and \eqref{eq:solutioniterative} represent small corrections to the linearized equation in Eq.~\eqref{eq:UGgravity} and its solution in Eq.~\eqref{eq:solution}. Therefore, the iterative approach, defined by Eq.~\eqref{eq:UGgravityn}, must converge to the solution of the exact nonlinear equation in Eq.~\eqref{eq:UGgravitygeneral}. Detailed study of the convergence properties of this solution is left as a topic of further work.

\section{\label{sec:TEGR}Obtaining TEGR from UG}

The starting point for obtaining TEGR from UG in Ref.~\cite{Partanen2025a} is the gauge-invariant Lagrangian density. In the present work, the gauge-invariant Lagrangian density is given in Eq.~\eqref{eq:LLL}. We also assume the equivalence principles of UG, given in Eq.~\eqref{eq:equivalence}. The formulation of TEGR is considered in the Weitzenböck gauge, where the teleparallel spin connection vanishes \cite{Aldrovandi2012,Bahamonde2023a}.

Since UG is formulated in the global Minkowski spacetime, its gravity gauge field has no relation to the metric or tetrad. Therefore, there is no equivalence transformation from UG to TEGR, where the fundamental field is the spacetime-dependent tetrad field, denoted by $\oset{\bullet}{e}{}_{\;\,\mu}^a$. However, it is found that the gauge-invariant Lagrangian density of UG in Eq.~\eqref{eq:LLL} reproduces the Lagrangian density of TEGR, if the following substitutions are made:
\begin{equation}
 I_\mathrm{g}^{a*}\mathcal{D}_\nu I_\mathrm{g}^a
 \longrightarrow-i\oset{\bullet}{e}_{a\nu},
 \label{eq:Ig_identityTEGR}
\end{equation}
\begin{equation}
 \delta_\mu^a\longrightarrow\oset{\bullet}{e}{}_{\;\,\mu}^a,
 \label{eq:TEGRtetrad}
\end{equation}
\begin{equation}
 \eta_{\mu\nu}\longrightarrow g_{\mu\nu}=\eta_{ab}\oset{\bullet}{e}\!{}_{\;\,\mu}^a\!\oset{\bullet}{e}\!{}_{\;\,\nu}^b,
 \label{eq:TEGRmetric}
\end{equation}
\begin{equation}
 d^4x\longrightarrow\sqrt{-\det(g_{\mu\nu})}d^4x.
 \label{eq:TEGRvolume}
\end{equation}
In TEGR, the Latin indices are raised and lowered by the Minkowski metric, and the Greek indices are raised and lowered by the spacetime-dependent metric in Eq.~\eqref{eq:TEGRmetric}. This fundamentally differs from UG, which uses the spacetime-independent tetrad, which is trivially $\delta_\mu^a$ when the Cartesian spacetime coordinates are assumed according to the present work.

The extension term of the Lagrangian density to account for the graviton--graviton interaction, given in Eq.~\eqref{eq:Lggint}, does not contribute to the Lagrangian density of TEGR, obtained in Ref.~\cite{Partanen2025a}. For this term, using Eq.~\eqref{eq:Ig_identityTEGR}, we obtain
\begin{align}
 \mathcal{L}_\mathrm{gg,int}=-i\sum_a T_\mathrm{g}^{a\nu}I_\mathrm{g}^{a*}\mathcal{D}_\nu I_\mathrm{g}^a
 \longrightarrow -T_\mathrm{g}^{a\nu}\oset{\bullet}{e}_{a\nu}
 =-T_\mathrm{g}{\!}^\nu_{\;\,\nu}=0.
 \label{eq:TEGRderivation}
\end{align}
The last equality of Eq.~\eqref{eq:TEGRderivation} gives zero, since the SEM tensor of the gravity gauge field in Eq.~\eqref{eq:Tg} is traceless. Therefore, the approach of obtaining TEGR from UG works the same way as studied in Ref.~\cite{Partanen2025a}. We point out that the relations in Eqs.~\eqref{eq:Ig_identityTEGR}--\eqref{eq:TEGRvolume} break the $4\times \mathrm{U(1)}$ gauge symmetry of UG, and TEGR is fundamentally different from the Minkowski spacetime formulation of UG.

\vspace{0.3cm}

\section{\label{sec:conclusion}Conclusion}
We have presented UG using a four-vector and tensor formalism. We have also introduced an extension of the original Minkowski spacetime formulation of UG to account for graviton--graviton interaction. This extension was obtained by adding the gauge-invariant SEM tensor of the gravity gauge field to the SEM tensors of the other fields in the Lagrangian density of UG. Therefore, the $4\times \mathrm{U(1)}$ gauge invariance of UG is preserved. The gauge-fixed Lagrangian density of UG also satisfies the global BRST invariance. The extension of UG presented in this work is necessary for correct description of gravitational interaction in problems involving interactions between gravitational fields, for example, in the propagation of gravitational waves in external gravitational potentials. Regarding the Feynman diagrams of UG, our extension introduces the triple-graviton vertex, whose implications to the renormalization of UG are left as a topic of further work. We have also discussed the relation between UG and TEGR.





\begin{acknowledgments}
This work has been funded by the Research Council of Finland under Contract No.~349971. We thank Holger Merlitz for discussions on the equivalence principle and the experimental tests of UG. We also thank Claudio Paganini for discussions on the experimental tests of UG.
\end{acknowledgments}

\end{document}